\documentclass[twocolumn,showpacs,showkeys,preprintnumbers,amsmath,amssymb]{revtex4}
\usepackage [dvips]{graphicx}

\usepackage [english]{babel}

\begin{document}

\title {Optical conductivity of metal nanofilms and
nanowires: The rectangular-box model}

\author {Valery P. Kurbatsky and  Valentin V. Pogosov\footnote{Corresponding author:
vpogosov@zntu.edu.ua}}

\address {Department of Micro- and Nanoelectronics,
Zaporozhye National Technical University, Zhukovsky Str. 64,
Zaporozhye 69063, Ukraine}

\date {\today}

\begin {abstract}
The conductivity tensor is introduced for the low-dimensional
electron systems. Within the particle-in-a-box model and the
diagonal response approximation, components of the conductivity
tensor for a quasi-homogeneous ultrathin metal film and wire are
calculated under the assumption   $d\cong \lambda_{\rm F}$ (where
$d$ is the characteristic small dimension of the system,
$\lambda_{\rm F}$ is the Fermi wavelength for bulk metal). We find
the transmittance of ultrathin films and compare these results with
available experimental data. The analytical estimations for the size
dependence of the Fermi level are presented, and the oscillations of
the Fermi energy in ultrathin films and wires are computed. Our
results demonstrate the strong size and frequency dependences of the
real and imaginary parts of the conductivity components in the
infrared range. A sharp distinction of the results for Au and Pb is
observed and explained by the difference in the relaxation time of
these metals.

\end {abstract}


\maketitle

\section {Introduction}

Thin-film materials, in particular, metal films are used widely in
modern technologies including electronics. As a rule, ultrathin
films are fragmentized (the island films) and consist of flat
islands connected with the thin threads-channels
\cite{Fed,Otero,Gloskovskii}.

Experimental techniques allow the optical characteristics in the
infrared range to be measured not only for thin films (see, for
example, \cite{4,5,7,9,10,11,12,13,14,16,17,17a}), but also for the
specifically grown nanorods-antennas of the micrometer length
\cite{Neubrech-2006,Muskens-2007,Neubrech-2008}.

In \cite{9}, the authors for the first time measured the infrared
conductivity of Pb ultrathin films. A decrease in the conductivity
of the films was explained by their granular structure.
Subsequently, Tu et al. \cite{11} measured the optical
characteristics of metal films at a temperature of 10 K and revealed
an anomalous optical transparency in the far-IR range. Pucci et al.
\cite{17} were the first to study the quantum size effects in the
transmission spectra of lead thin films by IR spectroscopy. It
should be noted that results of measurements have been usually
interpreted by experimenters in the framework of the modified Drude
theory. A theoretical analysis of optical properties of ultrathin
films and wires is necessary, in particular, for the diagnostics of
the nanostructure materials \cite{POGBOOK-2006} in order to use them
in micro- and nanoelectronics \cite{2}.

The important feature of the metal 1D- and 2D systems, films and
wires, is an anisotropy of their electrical and optical properties
caused by the size quantization. For this reason, the conductivity
of the low-dimensional systems is represented by a tensor
$\sigma_{\alpha\beta}(\textbf{q},\omega)$  which, in particular,
determines the optical absorption. The dissipation of energy of the
plane monochromatic electromagnetic wave with the frequency $\omega$
and the wave vector  $\textbf{q}$ in unit volume per unit time for a
nonmagnetic material is
$$
Q(\textbf{q},\omega)=\frac{1}{4}\sum\limits_{\alpha,\beta}
\left\{\sigma^{*}_{\alpha\beta}(\textbf{q},\omega)+
\sigma_{\beta\alpha}(\textbf{q},\omega)\right\}E_{\alpha}E^{*}_{\beta},
$$
where $E_{\alpha,\beta}$  are the components of the electric field
\cite{Landau}. However, the only value directly measurable for an
ultrathin film in IR range is the transmittance.

The purpose of this work is to calculate components of the
conductivity tensor for quasi-homogeneous ultrathin metal films and
wires provided the condition  $d\cong \lambda_{\rm F}$  is
satisfied. We use the Wood and Ashcroft approach \cite{18} adapted
to this case. The main advancement is the procedure of the accurate
determination of the Fermi level for a film and a wire of such a
thickness with taking into account the size oscillations. The
transmittance of the ultrathin films is also calculated in order to
compare theoretical results with experimental data \cite{4,5}.

To the best of our knowledge, the oscillatory behavior of the Fermi
energy in confined (2D) electron gas was studied for the first time
by Sandomirskii \cite{Sandom}. Later calculations were performed for
thin films, spheres  and wires
\cite{Schulte,Ek,8.,9.,10.,11.,12.,13.,25,hor,Han}. In the present
paper, in order to describe the Fermi energy behavior in
low-dimensional metallic systems we use an elementary one-particle
analytical approach \cite{25}.

\section {Conductivity tensor}

A film of thickness $L$ (or a wire of radius $\rho_{0}$) comparable
in magnitude to the Fermi wavelength of an electron in an infinite
metal ($\lambda_{\rm F}^{0}\approx 0.5$ nm) will be referred to as
the ultrathin film or wire (see Fig. 1). The longitudinal sizes of
the sample are assumed to be considerably larger than the film
thickness: $L\ll a,b$ (or $\rho_{0}\ll {\mathcal L}$ for wire),
which leads to the pronounced quantization of the transverse
component of the electron momentum. This results in the formation of
subbands, i.e., groups of energy levels corresponding to the same
value of the transverse momentum component.

A response of an electron gas to the electromagnetic field ${\rm
\textbf{E}}={\rm \textbf{E}}_{0}\exp[i_{0}(\textbf{q}
\textbf{r}-\omega t)]$ may be determined in a linear approximation
by the density matrix technique.

For the induced current one can obtain \cite{18}:
\begin{widetext}
\begin{multline}
\hat{\textbf{j}}(\textbf{k},\omega)=\frac{i_{0}e^{2}} {\Omega
m_{e}\omega}\left\{\textbf{E}_{0}\sum_{i}f_{i}\langle
i|e^{i_{0}(\textbf{q}-\textbf{k})\textbf{r}} |i \rangle \right.
\\
+ \left. \frac{1}{m_{e}}
\sum_{ij}\frac{f_{i}-f_{j}}{\varepsilon_{ij}- \hbar w}\left(\langle
j|e^{-i_{0}\textbf{k}\textbf{r}}\hat{\textbf{p}}|i\rangle
-\frac{1}{2}\hbar \textbf{k}\langle
j|e^{-i_{0}\textbf{k}\textbf{r}}|i\rangle \right) \left(\langle
i|e^{i_{0}\textbf{q}\textbf{r}}\textbf{E}_{0}
\hat{\textbf{p}}|j\rangle +\frac{1}{2}\hbar
\textbf{q}\textbf{E}_{0}\langle
i|e^{i_{0}\textbf{q}\textbf{r}}|j\rangle \right)\right\},
\label{(10cor)}
\end{multline}
\end{widetext}
where $i_{0}=\sqrt{-1}$; $|i\rangle$,  $|j\rangle$ are the wave
functions of the initial and final electron states corresponding to
energies $\varepsilon_{i}$ and $\varepsilon_{j}$;
$\varepsilon_{ij}=\varepsilon_{i}-\varepsilon_{j}$; $f_{j}$ and
$f_{j}$ are occupation factors; $\Omega$ is the volume of sample,
$m_{e}$ is the electron mass, $-e$ is the electron charge, ${\rm
\hat{\textbf{p}}}$ is the momentum operator.
\begin{figure}[!t!b]
\centering
\includegraphics [width=.4\textwidth] {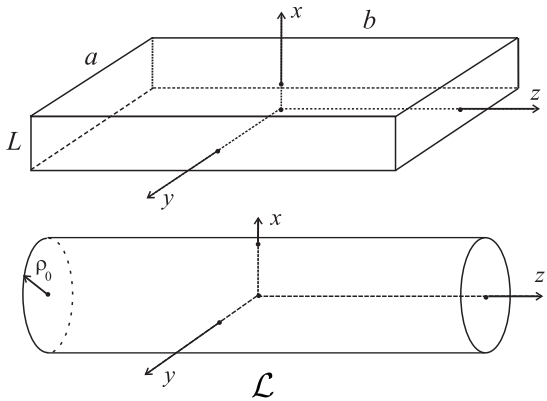}
\caption{Choice of coordinates.} \label{KP-fig1}
\end{figure}

We describe conductivity electrons in metal films and wires within
the framework of the particle-in-a-box model. In many cases, this
model turns out to be quite productive for metals with high
conductivity \cite{27}. In the case of the low-dimensional electron
systems, a potential box model includes the key feature of these
systems -- a confinement of electrons inside a region with certain
small dimension. The model remains applicable up to a certain
critical size when conductivity vanishes. As for the depth of a box,
this parameter remains important as long as electron emitting is
permitted.

In general, the model of electron gas in a rectangular potential box
is a good initial approximation, while various details
(particularities of structure, impurities etc.) can be taken into
account by introducing corresponding corrections.

An electron system in low-dimensional structures is anisotropic, and
its characteristics can be expressed in tensor form. The tensor
origin of the conductivity becomes obvious after converting the
expression (\ref{(10cor)}) into the form
$$
j_{\alpha}(\textbf{k},\omega)=\sum_{\beta}\sigma_{\alpha\beta}
(\textbf{k},\textbf{q},\omega)E_{\beta}(\textbf{q},\omega),
$$
where $\alpha,\beta=x,y,z$ and $\sigma_{\alpha\beta}$  is the
conductivity tensor.

It is not difficult to demonstrate that the conductivity tensor is
proportional to
$\delta_{\alpha\beta}\delta_{\textbf{k},\textbf{q}}$, where
$\delta_{\alpha\beta}$  or
$\delta_{\textbf{k},\textbf{q}}=\{1,\,\,\textbf{k}=\textbf{q};\,0,\,\,\textbf{k}\neq\textbf{q}\}$
is Kronecher's symbol, for macroscopic samples with the wave
functions of the kind $\Omega^{-1/2}\exp{(-\textbf{pr}/\hbar)}$.
This implies that all the Fourier components of the current, except
the one with $\textbf{k}=\textbf{q}$, are equal to zero. Of course,
this is not the case for ultrathin films and wires, but the
component with $\textbf{k}=\textbf{q}$ is still dominating. At the
first step, known as the diagonal response approximation, this
component only is taken into account.  We then find
\begin{widetext}
\begin{multline}
\sigma_{\alpha\beta}(\textbf{q},\textbf{q},\omega) =\frac{{i_{0}e^2
N}}{{m_e \omega\Omega}}\delta_{\alpha\beta}+
\frac{{i_{0}e^2}}{{m_e^2\omega\Omega}}\sum\limits_{i,j}{\frac{{f_i-f_j}}{{\varepsilon_{ij}
-\hbar\omega}}}\left({\left\langle
j\right|e^{-i_{0}\textbf{q}\textbf{r}} \hat p_\alpha
\left|i\right\rangle -\frac{1}{2}\hbar q_\alpha\left\langle
j\right|e^{-i_{0}\textbf{q}\textbf{r}}\left|i\right\rangle}\right)
\\
\times\left({\left\langle i\right|e^{i_{0}\textbf{q}\textbf{r}}\hat
p_\beta \left|j\right\rangle + \frac{1}{2}\hbar q_\beta\left\langle
i \right|e^{i_{0}\textbf{q}\textbf{r}}\left| j \right\rangle }
\right)\equiv \sigma_{\alpha\beta}(\textbf{q},\omega).
\label{KKBP-10}
\end{multline}
\end{widetext}
Here the relation $\sum_{i}f_{i}=N$ is used with $N$ equal to the
number of the conductivity electrons.

Over infrared region, the condition  $qL,\,q\rho_{0}\ll 1$  is
satisfied allowing us to express the conductivity tensor in terms of
the according small value.

\section{FILM}

It is assumed that the conduction electrons of the film are located
in a rectangular potential box $V(\textbf{r})$ with a depth
$U_{0}<0$, so that the box shape reproduces the film shape (see Fig.
1), and
\begin{equation}
|U_{\rm 0}|=\varepsilon_{\rm F}^{0}+W_{0 }, \quad \varepsilon_{\rm
F}^{0}=\frac{\hbar^{2}}{2m}(3\pi^{2}\bar{n})^{2/3}. \label{Us}
\end{equation}
Here $W_{\rm 0},\,\varepsilon_{\rm F}^{0} $,  and $\bar{n}$ are the
electron work function, the Fermi energy and  the electron
concentration  for a bulk metal, respectively.

The unperturbed states of the film are described by the wave
functions
\begin{equation}
\Psi_{mnp}(x,y,z)=\frac{1}{\sqrt{ab}}\psi_{m}(x)e^{2\pi
ni_{0}y/a}e^{2\pi pi_{0}z/b}, \label{(15)}
\end{equation}
where $n,\,p=\pm1,\pm2,\ldots$ and $m = +1, +2, \ldots$. The
subscript $m$ numbers the subbands. The wave functions $\psi_{m}(x)$
are represented in the following form:

for even values of $m$,
\begin{multline}
\psi_{m}(x) = \left\{
\begin{array}{ll}
C_{m} \sin k_{xm} x,\,\,\, - L/2 < x < L/2, \\
(- 1)^{(m/2) + 1} B_{m}e^{-\kappa_{m}x} ,\,\, x > L/2, \\
(- 1)^{m/2} B_{m}e^{\kappa_{m}x} ,\,\, \,\,\,\,\,\,\,\,x <  - L/2, \\
\end{array}
\right. \label{(16)}
\end{multline}

and for odd values of $m$,
\begin{multline}
\psi_{m}(x) = \left\{
\begin{array}{ll}
C_{m} \cos k_{xm}x,\,\,\, - L/2 < x < L/2, \\
(-1)^{(m - 1)/2} B_{m} e^{-\kappa_{m}x},\,\,x > L/2, \\
(-1)^{(m - 1)/2} B_{m} e^{\kappa_{m} x},\,\,x < -L/2, \\
\end{array}
\right. \label{(17)}
\end{multline}
$$
C_{m}=\sqrt{\frac{2\kappa_{m}}{2+\kappa_{m}L}},\,\,\,\,\,
B_{m}=C_{m} \frac{k_{xm}}{k_{0}}e^{\kappa_{m} L/2}.
$$
Here, $C_{m}$ is the normalization factor, $k_{xm}$ are the roots of
the equation
\begin{equation}
k_{xm}L=-2\arcsin(k_{xm}/k_{0})+\pi m, \label{(18)}
\end{equation}
where $\kappa_{m}=\sqrt{k_{0}^{2}-k_{xm}^{2}}$ and $\hbar
k_{0}=\sqrt{2m_{e}|U_{0}|}$ (see Ref. \cite{25}).

In this section, we focus on optical transitions between subbands
accompanied by changing the transverse component of the electron
wave vector $k_{xm}$. These transitions contribute to the
$\sigma_{xx}$ component of the conductivity tensor. Since $qL\ll 1$,
we have in zero approximation
\begin{multline}
\sigma_{xx}=\frac{{i_{0}e^2 }}{{m_e \omega\Omega}}
\\
\times\left(N+
\frac{1}{m_e}\sum\limits_{i,j}{\frac{{f_i-f_j}}{{\varepsilon_{ij}
-\hbar\omega}}}\left|\langle j|e^{-i_{0}(q_{y}y+q_{z}z)} \hat p_x
|i\rangle \right|^{2}\right). \label{KKBP-10cor}
\end{multline}
Dividing by $\varepsilon_{ij} -\hbar\omega$ in the sum and
interchanging $i$ and $j$ for the second term appeared after this
dividing, expression (\ref{KKBP-10cor}) can be transformed into
\begin{multline}
\sigma_{xx}=\frac{{i_{0}e^2 }}{{m_e \omega\Omega}}
\\
\times\left(N+
\frac{2}{m_e}\sum\limits_{i,j}{\frac{{f_i\varepsilon_{ij}}}{{\varepsilon_{ij}^{2}
-\hbar^{2}\omega^{2}}}}\left|\langle j|e^{-i_{0}(q_{y}y+q_{z}z)}
\hat p_x |i\rangle \right|^{2}\right). \label{KKBP-10cor1111}
\end{multline}

Since
$$
\langle j|e^{-i_{0}(q_{y}y+q_{z}z)} \hat p_x |i\rangle = \langle
m'|\hat p_x |m\rangle
\delta_{q_{y},k_{yn}-k_{yn'}}\delta_{q_{z},k_{zp}-k_{zp'}},
$$
and in view of the fact that  $|k_{xm}-k_{xm'}|\gg q$, further
simplifications are possible:
\begin{multline}
\sigma_{xx}\approx \frac{{i_{0}e^2 }}{{m_e \omega\Omega}}
\\
\times\left(N+ \frac{2}{m_e}\sum\limits
_{\substack {m,m'\\
n,p}}\frac{f_{mnp}\,\varepsilon_{mm'}}
{\varepsilon_{mm'}^{2}-\hbar^{2}\omega^{2}} |\langle m'|\hat p_x
|m\rangle|^{2}\right). \label{KKBP-10cor1112}
\end{multline}
Here the occupation factor is approximated by the step function
$f_{mnp}=\theta (\varepsilon_{\rm F}-\varepsilon_{mnp})$, where
$\varepsilon_{\rm F}$ is the Fermi energy for nanofilm,
$\varepsilon_{mm'}=\hbar^{2}(k_{xm}^{2}-k_{xm'}^{2})/2m_{e}$. Using
Thomas-Reiche-Kuhn sum rule (see Ref. \cite{18}), we rewrite
(\ref{KKBP-10cor1112}) as
\begin{equation}
\sigma_{xx}= \frac{{2i_{0}e^2 \hbar^{2}\omega}}{{m_e^{2} \Omega}}
\sum\limits
_{\substack {m,m'\\
n,p}}\frac{f_{mnp}|\langle m'|\hat p_x |m\rangle|^{2}}
{\varepsilon_{mm'}(\varepsilon_{mm'}^{2}-\hbar^{2}\omega^{2})},
\label{KKBP-10cor1113}
\end{equation}
and then one can obtain corresponding component of the dielectric
tensor
\begin{equation}
\epsilon_{xx}=1+\frac{4\pi i_{0}}{\omega}\sigma_{xx}. \label{EPS}
\end{equation}

The matrix elements of the momentum projection operator
$\hat{p}_{x}=i\hbar \partial/\partial x$ from (\ref{(15)}) --
(\ref{(17)}) are
\begin{multline}
|\langle m'|\hat{p}_{x}|m\rangle|^{2}=\Big\{1-(-1)^{m+m'}\Big\}
\\
\times\frac{8\hbar^{2}k_{xm}^{2}k_{xm'}^{2}\kappa_{m}\kappa_{m'}}
{(k_{xm'}^{2}-k_{xm}^{2})^{2}(2+\kappa_{m}L)(2+\kappa_{m'}L)}.
\label{(19)}
\end{multline}

The broadening is introduced in a manner proposed by Mermin
\cite{26}. As a result of this procedure the tensor components
$\sigma_{xx}$ and $\epsilon_{xx}$ get both real and imaginary parts:
\begin{equation}
{\rm Re}\,\sigma_{xx} = \Big(\frac{4}{L} \Big)^{3} \frac{a_{0}
\gamma^{2}}{\pi} \Big(\frac{1}{\hbar}\frac{e^{2}}{2a_{0}}\Big)
H_{(+)}, \label{F9}
\end{equation}
\begin{equation}
{\rm Im}\,\sigma_{xx} = - \Big(\frac{4}{L} \Big)^{3} \frac{a_{0}
k_{\omega}^{2} }{\pi }\Big(\frac{1}{\hbar}\frac{e^{2}}{2a_{0}}\Big)
H_{(-)}, \label{F10}
\end{equation}
\begin{equation}
{\rm Re}\,\epsilon_{xx}=1+\Big(\frac{4}{L}\Big)^{4}\frac{L}
{a_{0}}H_{(-)}, \,\,{\rm Im}\,
\epsilon_{xx}=\Big(\frac{4}{L}\Big)^{4}\frac{L\gamma^{2}}{a_{0}k_{\omega}^{2}}H_{(+)},\label{(24a)}
\end{equation}
\begin{widetext}
where
\begin{equation}
H_{(\mp)}=\sum_{m=1}^{m_{\rm F}}\sum_{m'=1}^{m_{\rm max}}
\{1-(-1)^{m+m'}\}\frac{L^{2}\kappa_{m}\kappa_{m'}k_{xm}^{2}k_{xm'}^{2}(k_{\rm
F}^{2}-k_{xm}^{2})\{(k_{xm'}^{2}-k_{xm}^{2})^{2}\mp
k_{\omega}^{4}\mp
\gamma^{4}\}}{(2+\kappa_{m}L)(2+\kappa_{m'}L)(k_{xm'}^{2}-k_{xm}^{2})^{3}\{\{(k_{xm'}^{2}-k_{xm}^{2})^{2}-
k_{\omega}^{4}+\gamma^{4}\}^{2}+4k_{\omega}^{4}\gamma^{4}\}}.
\label{(25)}
\end{equation}
\end{widetext}
Here $\gamma=\sqrt{2m_{e}/\hbar \tau}$, $\tau$ is the relaxation
time, $\hbar k_{\omega}=\sqrt{2m_{e}\hbar\omega}$,  $a_{0}$  is the
Bohr radius and
\begin{equation}
m_{\rm F}=\left[\frac{Lk_{\rm
F}}{\pi}+\frac{2}{\pi}\arcsin\Big(\frac{k_{\rm
F}}{k_{0}}\Big)\right],\, m_{\rm
max}=\left[\frac{Lk_{0}}{\pi}\right]+1. \label{(21)}
\end{equation}
Square brackets in (\ref{(21)}) and in the text below indicate the
integer number. Instead of the summation over $n$ and $p$ in
(\ref{KKBP-10cor1113}), (\ref{EPS}) we perform integration.

In order to use Eqs. (\ref{F9}) -- (\ref{(21)}) in calculations, it
is necessary to supplement them by the relation determining the
Fermi energy of film \cite{25}
\begin{equation}
k_{\rm F}^{2} =\frac{1}{m_{\rm F}}\left(2\pi
\bar{n}L+\sum_{m=1}^{m_{\rm F}}k_{xm}^{2} \right).
\label{Kfermifilm}
\end{equation}
The relation (\ref{Kfermifilm}) together with Eqs. (\ref{(18)}) and
(\ref{(21)}) describes the size-dependent Fermi level in ultrathin
films.

In the case of a film, transmittance is a quantity, which is
directly measurable
\begin{equation}
{\rm TR}= I/I_{0}, \label{(30a)}
\end{equation}
where $I_{0}$ and $I$ are intensities of a wave at surfaces $x=-L/2$
and $x=L/2$, respectively.

For a film of thickness $L$ the transmittance may be estimated as:
\begin{equation}
{\rm TR}=\exp\{-\eta(\omega,L) L\}, \label{(30)}
\end{equation}
where the absorption coefficient $\eta$ should be calculated by
using (\ref{(24a)}) and the formula
\begin{equation}
\eta=\frac{2\omega}{c} {\rm Im} \sqrt{\epsilon(\omega,L)}.
\label{(28)}
\end{equation}

\section{wire}

The simplest model for an ultrathin wire (see Fig. 1) is to consider
it as a cylindrical potential well $V(\rho,z)$ of infinite depth.
The length of the well ${\mathcal L}$ is assumed to be much larger
than its radius $\rho _0$. The conductivity electrons are described
by the wave functions of the kind
\begin{equation}
\psi _{mnp} \left( {\rho ,\varphi ,z} \right) = R_{mn} \left( \rho
\right)\Phi _m \left( \varphi  \right){\rm Z}_p \left( z
\right).\label{PPP}
\end{equation}

The function
\begin{equation}
{\rm Z}_p \left( z \right) = \frac{1}{{\sqrt {\mathcal L} }}e
^{i_{0}k_{zp}z} \label{KKBP-2}
\end{equation}
corresponds to the longitudinal motion of an electron. The subscript
$p$ numbers values of $z-$component of its wave vector. The angle
part of the wave function
\begin{equation}
\Phi _m \left( \varphi  \right) = \frac{1}{{\sqrt {2\pi }
}}\,e^{i_0m\varphi}   \label{KKBP-3}
\end{equation}
has to satisfy the periodicity condition
\begin{equation}
\Phi_{m}\left(\varphi+2\pi\right)=\Phi_{m}\left(\varphi\right),
\label{KKBP-4}
\end{equation}
from which follows the eigenvalues spectrum $m = 0, \pm 1,$ $ \pm
2,\ldots$.

The radial dependence of the wave function is described by the
Bessel functions of an integer order
\begin{equation}
R_{mn} \left( \rho  \right) = C_{mn}I_m \left(k_{mn} \rho \right),
\label{KKBP-5}
\end{equation}
where
\begin{equation}
C_{mn} =\frac{\sqrt{2}}{{\rho _0 \left| {I'_m \left( {k_{mn} \rho _0
} \right)} \right|}}. \label{KKBP-5a}
\end{equation}
Here $k_{mn}  = a_{mn} /\rho _0$, where $a_{mn}$ are positive roots
of the Bessel function of the $m$-th order $I_m(\xi)$, $n = 1,
2,\ldots$. The prime marks a derivative with respect to $\xi$.

In the next Section, for the case of a wire, we obtain the relation
similar to Eq. (\ref{Kfermifilm}).

\subsection{The Fermi energy}

We start from the expression for the energy of an electron
\begin{equation}
\varepsilon_{mnp}
=\frac{\hbar^{2}}{2m_{e}}\left(k_{mn}^{2}+k_{zp}^{2}
\right),\label{YU}
\end{equation}
where $k_{mn}$ and $k_{zp}$ are the eigenvalues of transverse and
longitudinal components of the electron wave vector, respectively
\cite{2}. The electron states in a wire correspond to points
($k_{mn},k_{zp}$) on the $k_{\perp}$ $k_{zp}$ half-plane
($k_{\perp}> 0$). Since the spectrum $k_{zp}$ is quasicontinuous
(${\mathcal L}\gg\rho_{0}$), these points form a system of straight
lines  $k_{\perp}=k_{mn}$. The occupied states distribute on
intercepts cut off by the semicircle of radius $k_{\rm F}$ (see Fig.
2). Density of the electron states at the intercepts is equal to
${\mathcal L}/\pi$.
\begin{figure}[!t!b!p]
\centering
\includegraphics [width=.4\textwidth] {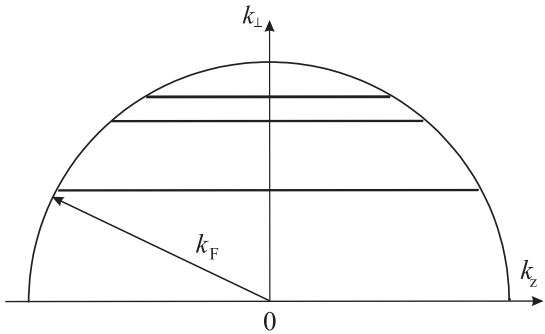}
\caption{Geometrical diagram of electron-state filling in quantum
wire.} \label{KP-fig2}
\end{figure}

The total number of the occupied states (equal to the number of the
conduction electrons in a wire) is
$$
N = 2\frac{{\mathcal L}}{\pi }\sum\limits_{m,n} {\sqrt {k_{\rm F}^2
- k_{mn}^2 } }.
$$
Taking into account that $N = \bar{n}\Omega$, we then obtain the
equation for the computation of the Fermi level $k_{\rm F}$ in an
ultrathin wire
\begin{equation}
\bar{n}=\frac{2}{{\pi^2\rho_0^2}}\sum\limits_{m,n}{\sqrt{k_{\rm F}^2
-  k_{mn}^2 }}. \label{KKBP-1.0}
\end{equation}
The electron concentration  $\bar{n}$ is assumed to be the same in a
wire and in a bulk metal. The summation should be performed over all
numbers $m$ and $n$ satisfying the condition
\begin{equation}
k_{mn}\leq k_{\rm F}. \label{KKBP-8a}
\end{equation}

The size dependence of the Fermi level in ultrathin films and wires
has an "oscillatory'' form. In order to determine a magnitude of the
variations, let us evaluate the averaged (smoothed) size dependence.
In this case, an averaging means a replacement of summation in Eqs.
(\ref{Kfermifilm}) and (\ref{KKBP-1.0}) by integration.

For a film, we use the Euler-MacLaurin summation formula
\cite{Korn}, in which it is enough to take into account the first
two terms. Allowing $m_{\rm F}$  to take any value (not only
integer, $m_{\rm F}\approx Lk_{\rm F}/\pi$) and neglecting the
corrections for a finite depth of a potential box, we obtain
\begin{equation}
k_{\rm F}/k_{\rm F}^{0}\approx 1+\pi/(4k_{\rm F}^{0}L),
\label{UUUUU}
\end{equation}
where  $k_{\rm F}^{0}$ is the Fermi wave number for a bulk metal.

In the case of a wire, it is hard to estimate directly the size
dependence of the Fermi level because it is impossible to express
explicitly the roots of the Bessel functions. However, the averaged
size dependence $k_{\rm F}(\rho_{0})$  can be obtained in an
indirect way.

Let us rewrite (\ref{KKBP-1.0}) as
$$
(k_{\rm F}^{0}\rho_{0})^{3}=6\sum\limits_{m,n}{\sqrt{(k_{\rm
F}^{0}\rho_{0})^{2} - a_{mn}^{2}}}.
$$
Here $k_{\rm F}^{0}=(3\pi^{2}\bar{n})^{1/3}$, and the relation
\begin{equation}
k_{mn}=a_{mn}/\rho_{0} \label{UUUUU2}
\end{equation}
was used. $a_{mn}$ are positive roots of the Bessel function of
order $m= 0, \pm 1, \pm 2, \ldots$ and $n = 1, 2, \ldots$. Assuming
the $a_{mn}=a(m,n)$ function to be continuous, we turn to the
integration:
\begin{equation}
(k_{\rm F}^{0}\rho_{0})^{3}=12\int\int\sqrt{(k_{\rm F}\rho_{0})^{2}
- a^{2}(m,n)}\,\,dm\,dn \label{UUUUU1}
\end{equation}
($m\geq 0$ now). Limits of the integration are determined by the
condition  $a(m,n)\leq k_{\rm F}\rho_{0}$.

The left-hand side of Eq. (\ref{UUUUU1}) tends to zero with
$\rho_{0}\rightarrow 0$. At the same time, the right-hand side tends
to zero only if $k_{\rm F}\rho_{0}\rightarrow a_{01}$. Hence, the
averaged size dependence is
$$
k_{\rm F}(\rho_{0}) \approx a_{01}/\rho_{0}
$$
for the small values $\rho_{0}$. For the large values $\rho_{0}$,
the integrated expression is $k_{\rm F}\rho_{0}$ and the area of the
region of integration  $\int\int dm\,dn$ is proportional to $k_{\rm
F}^{2}\rho_{0}^{2}$. Then, comparing the left-hand and right-hand
sides, we find that $k_{\rm F}(\rho_{0})\rightarrow$ const  with
$\rho_{0} \rightarrow \infty$. Accepting the constant to be $k_{\rm
F}^{0}$, we finally obtain
$$
k_{\rm F}/k_{\rm F}^{0}=  1  + a_{01}/(k_{\rm F}^{0}\rho_{0}),
$$
where $a_{01}\approx 2.4048$.

In the next subsection, to calculate conductivity components, we use
the size-dependent Fermi energy $\varepsilon_{\rm F}$ found from the
exact expression (\ref{KKBP-1.0}).

\subsection {Components of the conductivity tensor}

Let us consider the case when a wave is directed normally to the
axis of a wire (see Fig. 1). The wave vector is then located in the
$x-y$ plane, i.e.,  $ q_{z} = 0$. Orientating the $x-$axis along the
wave propagation, we get  $ q_{y} = 0$,  $\textbf{q}\textbf{r} =
q_{x} x \simeq \rho _0 /\lambda \ll 1$, and  $e^{\pm
i_{0}\textbf{q}\textbf{r}} \approx 1 \pm i_{0}q_{x}x$.

In zero order of the expansion  $\sigma_{\alpha\beta}$ in terms of
$\rho_{0}/\lambda$, the expression (\ref{KKBP-10}) takes the form
\begin{equation}
\sigma _{\alpha \beta }  = \frac{{i_{0}e^2 N}}{{m_e \omega \Omega
}}\delta _{\alpha \beta }  + \frac{{i_{0}e^2 }}{{m_e^2 \omega \Omega
}}\sum\limits_{i,j} {\frac{{f_i  - f_j }}{{\varepsilon _{ij}  -
\hbar \omega }}} \left\langle j \right|\hat p_\alpha  \left| i
\right\rangle \left\langle i \right|\hat p_\beta  \left| j
\right\rangle.\label{KKBP-10a}
\end{equation}
Following the procedure, which led us to Eq. (\ref{KKBP-10cor1111}),
we have
\begin{multline}
\sigma _{\alpha \beta }  = \frac{{i_{0}e^2 N}}{{m_e \omega \Omega
}}\delta _{\alpha \beta }  + \frac{{i_{0}e^2 }}{{m_e^2 \omega \Omega
}}\sum\limits_{i,j}f_i
\\
\times\left({\frac{{\left\langle j \right|\hat p_\alpha  \left| i
\right\rangle \left\langle i \right|\hat p_\beta \left| j
\right\rangle}}{{\varepsilon _{ij} - \hbar \omega }}}
+{\frac{{\left\langle j \right|\hat p_\alpha \left| i
\right\rangle^{*} \left\langle i \right|\hat p_\beta \left| j
\right\rangle}^{*}}{{\varepsilon _{ij} + \hbar \omega
}}}\right).\label{KKBP-11}
\end{multline}

Using (\ref{KKBP-2}) -- (\ref{KKBP-5a}), after rather cumbersome
transformations (see Appendix A), the matrix elements of various
projections of the momentum operator can be written as
\begin{multline}
\left\langle j \right|\hat p_\alpha \left| i \right\rangle = \left\{
\begin{array}{ll}
\hbar k_{zp} \delta _{ij},\,\,\, \alpha =z; \\
-\frac{i_{0}\hbar}{2}\delta_{pp'}k_{mn}C_{mn}\mathcal{G}_{(-)},\,\, \alpha =x; \\
\frac{\hbar}{2}\delta_{pp'}k_{mn}C_{mn}\mathcal{G}_{(+)},\,\, \alpha =y; \\
\end{array}
\right. \label{KKBP-12}
\end{multline}
\begin{equation}
\mathcal{G}_{(\mp)}= \delta_{m-1,m'}\mathcal{J}_{(-)}\mp
\delta_{m+1,m'}\mathcal{J}_{(+)}, \label{KKBP-13}
\end{equation}
$$ \mathcal{J}_{(\mp)}= C_{m\mp
1,n'}\int\limits_0^{\rho_0}I_{m\mp 1}(k_{m\mp 1,n'}\rho)I_{m\mp
1}(k_{mn}\rho)\rho d\rho.
$$
Because of a specific form of $\left\langle j \right|\hat p_z \left|
i \right\rangle$ the sum in Eq. (\ref{KKBP-10a}) becomes zero if
$\alpha=z$ or $\beta =z$. Hence,
\begin{equation}
 \sigma
_{xz,zx,yz,zy}= 0,\,\,\,\,\, \sigma _{zz}  = \frac{i_{0}e^2
\bar{n}}{m_e \omega}. \label{KKBP-15}
\end{equation}
For other diagonal components, the expression (\ref{KKBP-11}) can be
easily transformed into
\begin{equation}
\sigma_{\alpha\alpha}=\frac{{i_{0}e^2 \bar{n}}}{{m_e \omega}}+
\frac{{2i_{0}e^2}}{{m_e^2\omega\Omega}}\sum\limits_{i,j}
{\frac{f_i\,\varepsilon_{ij}}{\varepsilon_{ij}^{2}
-\hbar^{2}\omega^{2}}}\left|\left\langle j|\hat p_\alpha
|i\right\rangle\right|^{2}, \label{KKBP-116}
\end{equation}
where the subscript  $\alpha=x, y$. After a substitution of the
matrix elements (\ref{KKBP-12}) into (\ref{KKBP-116}), we find
\begin{widetext}
\begin{equation}
\sigma_{xx,yy}=\frac{i_{0}e^{2}\bar{n}}{m_{e}\omega}+\frac{i_{0}e^{2}}{m_{e}\omega\Omega}
\sum\limits_{\substack {m,n\\
p,n'}}f_{mnp}k_{mn}^{2}C_{mn}^{2}\left\{ \frac{(k_{mn}^2 -
k_{m-1,n'}^2)\mathcal{J}_{(-)}^{2}}{(k_{mn}^2 - k_{m-1,n'}^2)^{2}-
k_\omega ^4} +\frac{(k_{mn}^2 -
k_{m+1,n'}^2)\mathcal{J}_{(+)}^{2}}{(k_{mn}^2 - k_{m+1,n'}^2)^{2}-
k_\omega ^4}\right\}, \label{KKBP-17}
\end{equation}
\end{widetext}
where
$$
f_{mnp}  = \left\{ \begin{array}{l}
 1,\,\,\,k_{mn}^2  + k_{zp}^2  < k_{\rm F}^2 , \\
 0,\,\,\,k_{mn}^2  + k_{zp}^2  > k_{\rm F}^2 . \\
 \end{array} \right.
$$

An expression for the non-diagonal components $\sigma_{xy}$  and
$\sigma_{yx}$ follows from (\ref{KKBP-11})
\begin{multline}
\sigma _{\alpha \beta } = \frac{{i_{0}e^2 }}{{m_e^2 \omega \Omega
}}\sum\limits_{i,\,j} f_i
\\\times
\left( {\frac{{\left\langle j \right|\hat{p}_x \left| i
\right\rangle \left\langle i \right|\hat{p}_y\left| j \right\rangle
}}{{\varepsilon _{ij}  \mp \hbar \omega }}} + {\frac{{\left\langle j
\right|\hat{p}_x\left| i \right\rangle ^* \left\langle i
\right|\hat{p}_y\left| j \right\rangle ^* }}{{\varepsilon _{ij}  \pm
\hbar \omega }}} \right). \label{KKBP-18}
\end{multline}
The upper sign corresponds to  $\alpha = x$, $\beta  = y$, and the
lower one to  $\alpha  = y$, $\beta  = x$.

The axis symmetry of the problem is reflected by the fact that in
(\ref{KKBP-10a}) and (\ref{KKBP-11}) the summation is performed over
positive $m$ and $m'$ as well as negative ones but the same in
absolute value. The analysis of the expressions (\ref{KKBP-12}) and
(\ref{KKBP-13}) based on the properties of the Bessel functions
\cite{Yanke}
$$
k_{(-m)n}=k_{mn},\,\,\,\  I_{-m}(\xi)=(-1)^{m}I_{m}(\xi)
$$
reveals a different behavior of the matrix elements when changing
together $ m \to - m $ and $ m' \to - m'$,
\begin{equation}
\left\langle j \right|\hat p_x \left| i \right\rangle \to
 - \left\langle j \right|\hat p_x \left| i \right\rangle,
\\
\left\langle j \right|\hat p_y \left| i \right\rangle \to
\left\langle j \right|\hat p_y \left| i \right\rangle.
\label{KKBP-19}
\end{equation}
This causes the terms in (\ref{KKBP-18}) to cancel pairwise, and we
then find
\begin{equation}
\sigma_{xy}=\sigma_{yx}=0. \label{KKBP-20}
\end{equation}

Thus, all non-diagonal components of the conductivity tensor vanish
in zero approximation of the expansion in terms of
$\rho_{0}/\lambda$. However, in linear approximation the result is
different. We take account that terms, which contain $\delta_{ij}$,
lead to the vanishing of the sum. Therefore, in this approximation,
components  $\sigma_{zx}$ and $\sigma_{zy}$ have a form
\begin{multline}
\sigma_{z\beta}=\frac{q_{x}e^{2}}{m_{e}^{2}\omega\Omega}\sum\limits_{i,j}f_i
\\\times
\left({\frac{{\left\langle j \right|x\hat p_z  \left| i
\right\rangle \left\langle i \right|\hat p_\beta \left| j
\right\rangle}}{{\varepsilon _{ij} - \hbar \omega }}}
+{\frac{{\left\langle j \right|x\hat p_z\left| i \right\rangle^{*}
\left\langle i \right|\hat p_\beta \left| j
\right\rangle}^{*}}{{\varepsilon _{ij} + \hbar \omega }}}\right)
,\label{KKBP-21}
\end{multline}
where $\beta=x, y$ and for the matrix elements see Appendix.

An analysis, similar to the one, which resulted in Eq.
(\ref{KKBP-19}), gives
\begin{multline}
\left\langle j \right|x\hat p_z \left| i \right\rangle \to
 - \left\langle j \right|x\hat p_z \left| i \right\rangle,
\\
\left\langle j \right|x\hat p_z \left| i \right\rangle \left\langle
i \right|\hat p_y \left| j \right\rangle  \to
 - \left\langle j \right|x\hat p_z \left| i \right\rangle \left\langle i \right|\hat p_y \left| j
 \right\rangle,
\\
\left\langle j \right|x\hat p_z \left| i \right\rangle \left\langle
i \right|\hat p_x \left| j \right\rangle  \to \left\langle j
\right|x\hat p_z \left| i \right\rangle \left\langle i \right|\hat
p_x \left| j \right\rangle .\label{KKBP-23}
\end{multline}
Hence, to linear order in $\rho_{0}/\lambda$  we have
$\sigma_{zy}=0$ but  $\sigma_{zx}\neq 0$. Using Eq. (\ref{KKBP-13}),
 the relation
$$
\left\langle j\right|x\hat p_{z}\left|i\right\rangle^{*}\left\langle
i\right|\hat p_{x}\left|j\right\rangle^{*}=-\left\langle
j\right|x\hat p_{z}\left|i\right\rangle\left\langle i\right|\hat
p_{x}\left|j\right\rangle
$$
and Eq. (\ref{KKBP-B3}), we derive
\begin{equation}
\sigma_{zx}=\frac{2i_{0}q_{x}e^{2}}{\hbar\Omega}\sum\limits
_{\substack {n,n'\\
m,p}}f_{mnp}k_{zp}C_{mn}^{2}(\mathcal{F}_{(-)} -\mathcal{F}_{(+)}),
 \label{KKBP-24}
\end{equation}
where
$$
\mathcal{F}_{(\mp)}=\frac{\mathcal{J}_{(\mp)}C_{m\mp 1,n'}
\int\limits_0^{\rho_0}I_{m\mp 1}\left(k_{m\mp 1,n'}\rho
\right)I_{m}\left(k_{mn}\rho\right)\rho^{2}d\rho}{(k_{mn}^{2}-k_{m\mp
1,n'}^{2})^{2}-k_{\omega}^{4}}.
$$

Dissipation is introduced by the substitution
$\omega\rightarrow\omega+i_{0}/\tau$ in expression for conductivity.
When $\tau=0$, the diagonal components of conductivity are
imaginary. Since the remaining components of the tensor vanish in
zero approximation, dissipation is also absent ($Q=0$). In general,
dissipation is small for optical frequencies in which we are
interested ($\omega \gg 1/\tau$).

Substituting $\omega\rightarrow\omega+i_{0}/\tau$ in
(\ref{KKBP-15}), after straightforward transformations we obtain the
Drude formula \cite{Ziman}
\begin{equation}
\sigma_{zz}(\omega)=\sigma(0)\frac{1+i_{0}\omega\tau}{1+\omega^{2}\tau^{2}},\label{KKBP-25}
\end{equation}
where  $\sigma(0)\equiv e^{2}\bar{n}\tau/m_{e}$  is the static
conductivity. Thus, the component $\sigma_{zz}(\omega)$  is
associated with the classical conductivity. Other diagonal
components (\ref{KKBP-116}) can be represented as
\begin{equation}
\sigma_{\alpha\alpha}=\sigma_{zz}\{1+S(\omega,\rho_{0},{\mathcal
L})\}, \label{KKBP-26}
\end{equation}
where
\begin{equation}
S\equiv\frac{2}{Nm_{e}} \sum\limits_{\scriptstyle
i,j}\frac{f_{i}\varepsilon_{ij}\left(\varepsilon_{ij}^{2}-
\hbar^{2}\omega^{2}+2\hbar^{2}\omega
i_{0}/\tau\right)}{(\varepsilon_{ij}^{2}
-\hbar^{2}\omega^{2})^{2}+4\hbar^{4}\omega^{2}/\tau^{2}}
\left|\langle j|\hat{p}_{\alpha}|i\rangle\right|^{2}
 \label{KKBP-27}
\end{equation}
and $\alpha=x,y$.

After interchanging subscripts $i$ and $j$, terms of the sum
(\ref{KKBP-27}) reverse their sign. As a result,
 $$
\sum_{\substack {i,j\\
\varepsilon_{i},\varepsilon_{j}<\varepsilon_{\rm F}}}
\frac{f_{i}\varepsilon_{ij}\left(\varepsilon_{ij}^{2}-
\hbar^{2}\omega^{2}+2\hbar^{2}\omega
i_{0}/\tau\right)}{(\varepsilon_{ij}^{2}
-\hbar^{2}\omega^{2})^{2}+4\hbar^{4}\omega^{2}/\tau^{2}}
\left|\langle j|\hat{p}_{\alpha}|i\rangle\right|^{2}=0,
$$                …
and
\begin{equation}
S= \frac{2}{Nm_{e}}\sum_{\substack {i,j\\
\varepsilon_{i}<\varepsilon_{\rm F}\\
\varepsilon_{j}>\varepsilon_{\rm F}}}
\frac{\varepsilon_{ij}\left(\varepsilon_{ij}^{2}-
\hbar^{2}\omega^{2}+2\hbar^{2}\omega
i_{0}/\tau\right)}{(\varepsilon_{ij}^{2}
-\hbar^{2}\omega^{2})^{2}+4\hbar^{4}\omega^{2}/\tau^{2}}
\left|\langle j|\hat{p}_{\alpha}|i\rangle\right|^{2}.
\label{KKBP-28}
\end{equation}
Here $\varepsilon_{ij}< 0$, i.e. only transitions coupled with
absorption participate in the conductivity. It is important to
remark that ${\rm Im}S<0$ for any frequency.  Since in the optical
region the real part of the component $\sigma_{zz}$ can be ignored
and its imaginary part is positive, it follows from (\ref{KKBP-26})
that ${\rm Re}\,\sigma_{xx,\,yy}>0$ and $Q>0$ over the whole region.

Let us compare in magnitude components of the conductivity tensor.
For Au, the frequency  $\hbar\omega=1$ eV, dissipation
$\hbar/\tau=0.02$ eV we find  $\sigma(0)=4.6\times 10^{17}$
s$^{-1}$,  $|\sigma_{zz}|\approx \sigma(0)/\omega\tau\approx
10^{16}$ s$^{-1}$. We use below the value
$e^{2}/2a_{0}\hbar=2.0\times 10^{16}$ s$^{-1}$ as a unit of the
conductivity. Then  $|\sigma_{zz}|\approx 0.5$.

We can now estimate, for example, height of peaks in ${\rm
Re}\sigma_{xx}$. We use relationships
$$
{\rm Re}\sigma_{xx}=-|\sigma_{zz}|{\rm Im}S
$$
and
$$
{\rm Im}S\approx-\frac{\tau}{\hbar Nm_{e}}\left|\langle
m+1,n'|\hat{p}_{x}|mn\rangle\right|^{2}\sum\limits_p 1
$$
(which may be obtained from (\ref{KKBP-26}) and (\ref{KKBP-27})
under condition that the peaks are well separated). Taking into
account that
\begin{equation}
\sum\limits_p 1=\frac{2{\mathcal
L}}{\pi}\sqrt{k_{F}^{2}-k_{mn}^{2}}\cong\frac{2{\mathcal
L}}{\pi}k_{\rm F}^{0}, \label{KKBP-2009}
\end{equation}
\begin{equation}
\left|\langle m+1,n'|\hat{p}_{x}|mn\rangle\right|^{2}\propto
k_{mn}^{2}\cong\frac{1}{4}\hbar^{2}{k_{\rm F}^{0}}^{2}
\label{KKBP-2010}
\end{equation}
and, using (\ref{Us}), we have
$$
{\rm Im}S\cong-\frac{3\hbar\tau}{2m_{e}\rho_{0}^{2}}.
$$
For $\tau=2.1\times 10^{-14}$ s$^{-1}$ (Au), $d=2\rho_{0}=2$ nm, we
find  ${\rm Im}S\cong -1$, ${\rm Re}\sigma_{xx}\cong 1$. In
macroscopic limit $\rho_{0}\rightarrow\infty$ we find that $ {\rm
Re}\,\sigma_{xx}=0$ and ${\rm Im}\,\sigma_{xx}={\rm
Im}\,\sigma_{zz}$, as we have expected.

Comparing (\ref{KKBP-11}) with (\ref{KKBP-21}), one can obtain
$|\sigma_{zx}/\sigma_{xx}|\cong q_{x}\rho_{0}$. For $\lambda=10^{3}$
nm, $d=2$ nm we have $|\sigma_{zx}/\sigma_{xx}|\cong10^{-2}$.

\section{Results and Discussion}

The difference between our approach and the theory \cite{18} is
associated with peculiarities of a electronic levels distribution in
films/wires of nanometers thickness, when the  $d\cong \lambda_{\rm
F}$ condition is satisfied.

In this case, a number of subbands, formed as a result of the size
quantization, is small, while the contribution of each of them in
the sum (\ref{(25)}) is significant. Opposite assumptions are made
in \cite{18}: a characteristic size $d$ is so large that the number
of subbands is much larger than 1. Then, the separation between
neighboring subbands (with numbers $m$ and $m+1$) is small, while
contributions of individual items in the sum \cite{18} are not
significant anymore, so the summation can be substituted by
integration, as usually done in the case of a quasi-continuous
distribution. Discreteness, coming from the size quantization,
manifests itself only weakly. The Fermi level in films and wires
with small thickness noticeably differs from the Fermi level of a
bulk metal (30\% difference for a wire of 1 nm diameter, see Fig.
3). In \cite{18}, the authors use the Fermi level of a bulk metal to
find a number of subbands, while we take into account the size
dependence of the Fermi level, when determining a number of
subbands. For a few nanometers thickness, these numbers are found to
be different and, because they are small, a noticeable divergence in
results is revealed.

Finally, theory \cite{18} was developed in order to apply it for an
isotropic composite medium. Therefore, from the very beginning, the
direction of an applied field was considered as a preferred one. In
our approach, anisotropy of the metallic 1D and 2D systems (wires
and films) is taken into consideration, their conductivity and
dielectric function are assumed to be tensors that allows a response
of wires and films to be determined for any orientation in an
external field.

Anisotropy as well as discreteness manifests itself much stronger in
systems with the small characteristic size  $d\cong \lambda_{\rm
F}$. Under the condition $d \gg \lambda_{\rm F}$, both our theory
and theory \cite{18} lead to the same results.

\subsection{The Fermi energy}

Fig. 3 demonstrates the size dependence of the Fermi energy for
films and wires of Au and Al computed from Eqs. (\ref{Kfermifilm})
and (\ref{KKBP-1.0}). The size dependences have an "oscillatory''
form. In contrast to the Fermi energy of a film \cite{25}, the size
variation of the Fermi energy of a wire seems to be random. Input
parameters for calculations were taken from Ref.  \cite{25}.
\begin{figure}[!t!b!p]
\centering
\includegraphics [width=.35\textwidth] {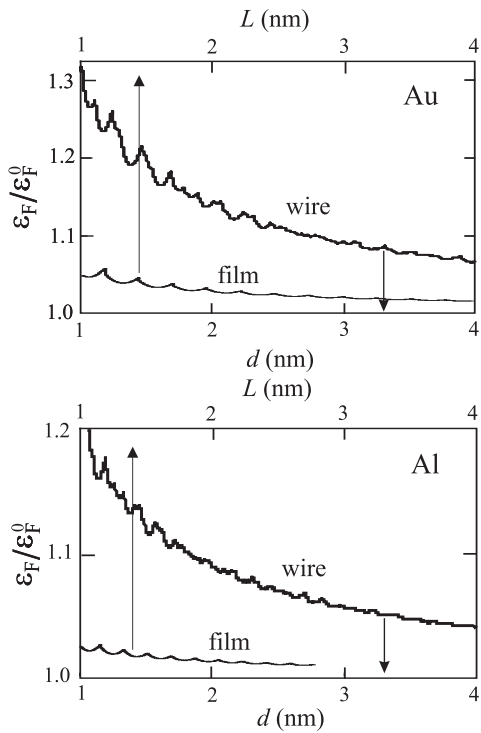}
\caption{Reduced size dependences of the Fermi energy of wires and
films vs diameter $d=2\rho_{0}$  and thickness $L$, respectively.}
\label{KP-fig3}
\end{figure}

In the case of a film, cusps on the size dependence (i.e., the jumps
of the derivative  $d\varepsilon_{\rm F}/dL$) are distributed nearly
regular with the approximately constant period  $\Delta
L\thickapprox\pi/k_{\rm F}^{0}$. The cusp on the size dependence of
a wire appears each time when the increasing radius $\rho_{0}$
reaches the value $\rho_{0 (m'n')}$ for which the condition
(\ref{KKBP-8a}) is satisfied by one more pair $(m',n')$:
$$
a_{m'n'}=k_{\rm F}\rho_{0(m'n')}.
$$

Distance between the neighboring cusps is
$$
\Delta d\approx 2\left(a_{m'n'}-a_{mn}\right)/k_{\rm F}^{0},
$$
where $a_{mn}$ is the root of the Bessel function closest to
$a_{m'n'}$  in value. Roots of the Bessel functions of different
orders mix up so that $\Delta d$ varies, at first sight, randomly
with change in size.

The oscillations of the Fermi energy in a wire of diameter $d$ and
in a film of thickness $L$ are similar in magnitude if $d\cong L$.
As in the case of a film, the "period'' $\Delta d$ and the amplitude
of the oscillations tend to zero with increasing diameter.

Characteristic properties of the size dependence of the Fermi energy
for various metal wires (and various metal films too) may be
explained exclusively by different value  $k_{\rm F}^{0}$. As
compared to the Au wire, for the Al wire, the scale $\Delta d$ of
the oscillations is finer, the amplitude of the oscillations and the
averaged value $\varepsilon_{\rm F}/\varepsilon_{\rm F}^{0}$ are
smaller.

\subsection{Film}

The specific feature of the optical characteristics of thin films is
the presence of peaks associated with the optical transitions
between the subbands. The size effect manifests itself in a change
of the number of peaks, their position, and the spacing between
them.

The positions of the peaks is determined by the approximate
expression $\hbar \omega_{mm'}\approx
\hbar\omega_{0}|{m'}^{2}-m^{2}|$, where $m$ and $m'$ are the numbers
of subbands between which the transition occurs and $\hbar
\omega_{0}\equiv \pi^{2}\hbar^{2}/(2m_{e}L^{2})=$ 0.34 [eV]$/L^{2}$
[nm$^{2}$]. The frequency range under consideration lies in the
infrared and visible spectral ranges. The lower limit of the range
($\hbar \omega_{12}$) corresponds to the beginning of the optical
transitions between the subbands. The upper limit of the frequency
range is the electron work function $W$ of the film. The estimates
can be made with the work function $W_{0}$  for infinite metals Au
and Ag.

The calculated real and imaginary parts of the conductivity
component $\sigma_{xx}$  for the Au films 2 and 6 nm thick are
presented in Fig. 4. For ultrathin films, the number of subbands
completely or partially occupied by electrons is small:  $m_{\rm
F}\approx 2L/\lambda^{0}_{\rm F}$. Therefore, the number of peaks is
small as well. For the film of the thickness $L=2$ nm, the peak at
$\hbar \omega_{12}\approx 0.25$ eV corresponding to the lower limit
of the frequency range is clearly seen. The peaks that represent the
transitions between the neighboring subbands with the numbers $m$
and $m'=m+1$ are located to the left of the maximum height peak
observed at the frequency  $\hbar\omega_{\rm max}\simeq
\hbar\omega_{0}(2m_{\rm F}+1)$. This frequency corresponds to the
transition between the subbands with the numbers $m=m_{\rm F}$ and
$m'=m_{\rm F}+1$. The spacing between any two neighboring peaks is
identical and approximately equal to  $2\hbar \omega_{c}$. As the
film thickness $L$ increases, all peaks shift toward the left, the
spacing between peaks decreases, and they begin to merge together.

The overlapping of the peaks becomes significant when the spacing
between them is equal to their width. The peak width is determined
by the dissipation mechanisms and is approximately equal to
$2\hbar/\tau$. Peaks  for the film of the thickness $L=2$ nm are
clearly distinguishable (see Fig. 4), but for the thickness $L=6$ nm
the peaks disappear completely. (It should be noted that the results
of our calculations appear to be weakly sensitive to a change in the
relaxation time $\tau$ within one order of magnitude.)
\begin {figure} [! t! b! p]
\centering
\includegraphics [width =.35\textwidth] {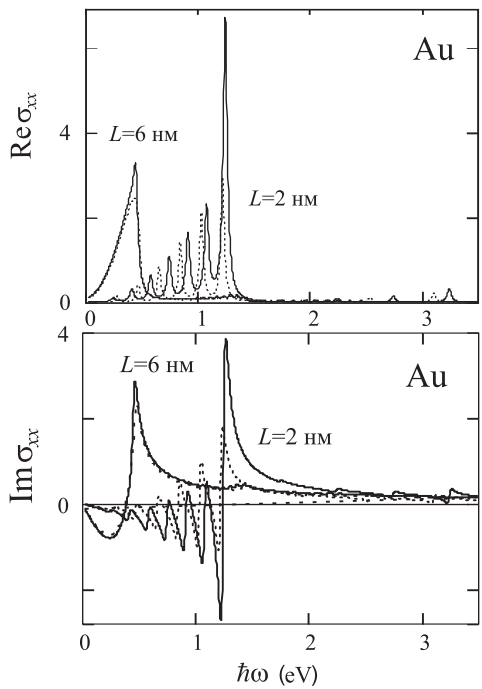}
\caption {Frequency dependences of the real and imaginary parts of
the film conductivity  component (in $e^{2}/2a_{0}\hbar$ units)
calculated by Eqs. (\ref{F9}) and (\ref{F10}) (solid lines). Dashed
curves  correspond to the results of calculations by formulas (68)
from work \cite{18}.} \label{KP-fig4}
\end {figure}

It can be seen from Fig. 4 that, as the film thickness decreases,
the discrepancy between the results of calculations by Eqs.
(\ref{F9}), (\ref{F10}) and by (68) from Ref. \cite{18}
 increases and becomes
substantial. This discrepancy is associated with the fact that
relationships (\ref{F9}) and (\ref{F10}) were derived with allowance
made for the dependence of the Fermi energy on the film thickness
$k_{\rm F}(L)$ and the exact calculation of the number $m_{\rm F}$
of occupied subbands. In  Ref. \cite{18}, the number $m_{\rm F}$ was
calculated by the procedure which gives an error $\pm 1$ for films
with thickness $L\simeq \lambda_{\rm F}$. This is an essential error
because the number of occupied subbands for such thickness is small
and ranges from 2 to 6.
\begin {figure} [! t! b! p]
\centering
\includegraphics [width =.4\textwidth]{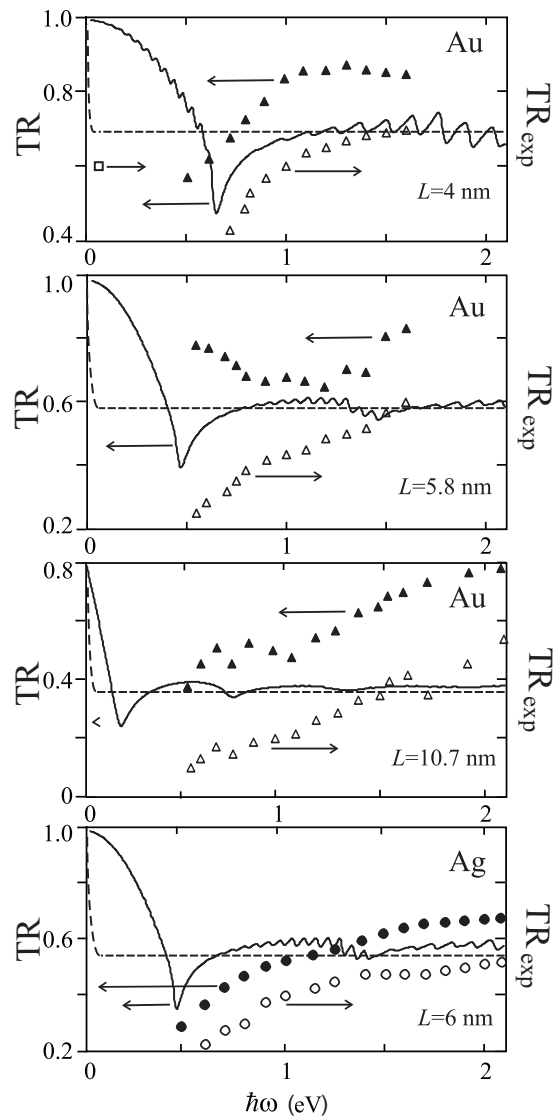}
\caption {Frequency dependences of the film transmittance calculated
by Eqs. (\ref{F9}) -- (\ref{(28)}) using $\epsilon_{xx}$ (solid
lines) and $\epsilon_{zz}$ (dashed lines, left hand scale). Opened
triangles \cite{5},  circles \cite{4}, and square \cite{11} indicate
the experimental data (right hand scale) for the Au and Ag films.
Solid triangles and circles indicate the recalculated experimental
data (left hand scale) in according to Eq. (\ref{REC}).}
\label{KP-fig5}
\end {figure}

The calculated frequency dependences of the transmittance for Au and
Ag thin films of different thicknesses are compared with the
experimental data in Fig. 5. When comparing the results of our
calculations with experimental data, it is necessary to take into
consideration that our definition of the transmittance is different
from that which is normally used by experimentalists.

In the transmittance (\ref{(30a)}) the value $I$ is the same in both
cases: This is an intensity of radiation, which comes out from the
film through the surface $x = L/2$. Experimentalists take $I_{0}$ as
intensity of radiation incident onto the surface of the film $x = -
L/2$. Of course, certain fraction of the radiation penetrates
inside, while the remaining part is reflected. We don't consider
reflection and assume $I_{0}$ to be intensity of radiation, which
\emph{comes into the film} through the surface $x = - L/2$. To make
a comparison with the theory, experimental values of the
transmittance $ {\rm TR}_{\rm exp}$ are recalculated by using the
formula
\begin{equation}
{\rm TR}=\frac{{\rm TR}_{\rm exp}}{1-{\rm R}}, \label{REC}
\end{equation}
where ${\rm R}$ is the value of reflection coefficient  obtained by
measuring under the same conditions as $ {\rm TR}_{\rm exp}$
\cite{5,4}. The results of the recalculation are also presented in
Fig. 5.

The absorbance $\eta$ is determined by the functions ${\rm Im}\,
\epsilon(\hbar\omega)$ and ${\rm Re}\, \epsilon(\hbar\omega)$
according to expression  (\ref{(28)}). The frequency dependences of
these functions exhibit a different behavior (Fig. 6).
\begin {figure} [! t! b! p]
\centering
\includegraphics [width =.35\textwidth] {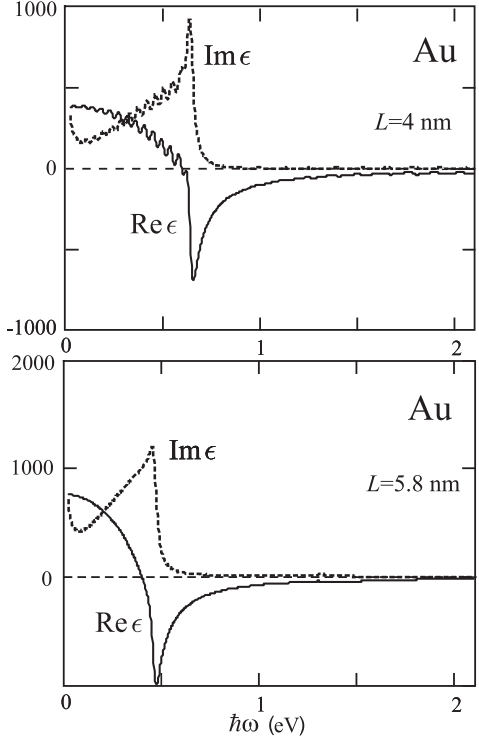}
\caption {Frequency dependences of the real and imaginary parts of
the film dielectric function  calculated by Eqs. (\ref{(24a)}).}
\label{KP-fig6}
\end {figure}

Unlike the function ${\rm Im}\, \epsilon(\hbar\omega)$, the function
${\rm Re}\, \epsilon(\hbar\omega)$ has not only pronounced resonance
maxima but also minima shown as inverted peaks. The height of both
peaks increases with an increase in the frequency, so that,
eventually, one of the inverted peaks intersects the abscissa axis,
and the function ${\rm Re}\, \epsilon(\hbar\omega)$ becomes negative
(in contrast to the function ${\rm Im}\, \epsilon(\hbar\omega)$ that
is always positive in sign). The minimum transmittance should be
identified with the minimum of the function ${\rm Re}\,
\epsilon(\hbar\omega)$, which is located in the vicinity of the
frequency $\hbar\omega_{\rm max}= \hbar\omega_{0}(2m_{\rm F}+1)$.

At frequencies $\hbar\omega>\hbar\omega_{\rm max}$, the absorbance
is determined only by the real part of the dielectric function:
$\eta\approx(2\omega/c)\sqrt{|{\rm Re}\, \epsilon|}$. It is easy to
check that the absorbance tends to a specific constant value with an
increase in the frequency. The transmittance TR (\ref{(30)}) is
characterized by the same tendency. This tendency can be clearly
seen in Fig. 5. The peaks associated with the transitions between
far subbands are clearly distinguished against the background of the
monotonic increase in the transmittance. In particular, the
transition $m_{\rm F}-3\rightarrow m_{\rm F}$ manifests itself at
$\hbar\omega\approx 1.8$ eV ($L=4$ nm).

To the zero order in $L/\lambda $, diagonal components of the
dielectric tensor only are not equal to zero. Solid lines in Fig. 5
represent transmittance computed by Eqs. (\ref{(30)}) and
(\ref{(28)}) using $\epsilon_{xx}$. The transmittance indicates a
change in a normal to surface component of an electric field of a
wave passed through a film. It is this component that causes optical
transitions between subbands formed by the size quantization. Dashed
lines represent transmittance calculated with using $\epsilon_{zz}$
($\epsilon_{zz}=\epsilon_{yy}$) which shows a weakening of a
parallel to surface component of an electric field. Such a
transmittance is observed at normal incidence of radiation onto a
film.

In the region $\hbar\omega>\hbar\omega_{\rm max}$, a mechanism of
dissipation (i.e. value of the relaxation time $\tau$) affects the
transmittance weakly. Dissipation manifests itself noticeably only
in vicinity of the minimum of transmittance. Thus, discrepancies
between theory and experiment at these frequencies could not be
explained by either a mechanism of dissipation or an orientation of
a film in field. A remarkable feature, as seen from Fig. 5, is a
noticeable exceeding of measured transmittance over computed one
(except of the last section, where there is a good agreement). This
implies that the discrepancy between the theory and experiment can
be attributed to a large non-homogeneity in thickness and especially
an absence of continuity, i.e. it can be explained by the presence
of regions of a substrate without coating.

Let's name the ratio of coating area to substrate area by the
coating coefficient $p$ and denote coating thickness as $L'$. The
mean thickness which is usually determined by experimentalists from
the mass of a film and substrate area is $L=pL'$. Transmittance of a
``holey'' film of thickness $L$ is
$$
{\rm TR}(L)=1-p+p{\rm TR}(L')=1-p+p{\rm TR}(L/p).
$$

Reducing $p$, it is possible to increase transmittance up to 1. For
example, transmittance of a film with thickness $L = 4$ nm and
coating coefficient $p = 0.45$ at the frequency $\hbar \omega =1$
 eV is equal to 0.75, i.e. discrepancy with the experimental value is
twice lowered (see Fig. 5).

In the frequency region $\hbar\omega < 0.5$ eV, there are
experimental data for transmittance of thin films of Au \cite{11}
and Pb \cite{11,17}. Unfortunately data for reflection are absent
hence the recalculation of experimental results like that
represented above is impossible. However, taking into account that
this recalculation leads to an increase of transmittance value, we
guess experimental data for Au  \cite{11} to be in agreement with
our calculations under assumption of a normal incidence of radiation
onto a film (Fig. 5). Moreover, the recalculation can change a type
of dependence on frequency for transmittance, i.e. a rising can be
replaced by a falling after the recalculation (see Fig. 5 for Au, $L
= 5.8$ nm). This could explain why within frequency interval
$(0.2,\,0.5)$ eV dependence of transmittance on frequency is
decreasing or absent at all according to our calculations whereas an
increasing is observed in experiment \cite{17}. As to value of
transmittance for Pb films, it is difficult to compare calculations
with experimental data because results of different experiments vary
essentially. Thus transmittance value for Pb film of 4 nm thickness
is given in \cite{11} as 0.79 at 0.05 eV but it is 0.12 only at the
close frequency 0.12 eV following \cite{17} (in addition,
transmittance diverges in value 1.5 times for different technologies
of film coating \cite{17}).

\subsection{Wire}

The frequency dependences of ${\rm Re}\,\sigma_{xx}$ and ${\rm
Im}\,\sigma_{xx}$ for the Au wire of diameter 1.6 nm are presented
in Fig. 7. For such a small diameter, the peaks corresponding to the
transitions between levels of the size quantization (subbands)
manifest itself clearly. In spite of the rather complete spectrum
$k_{mn}$, position of the peaks is well predicted.
\begin{figure}[!t!b!p]
\centering
\includegraphics [width=.4\textwidth] {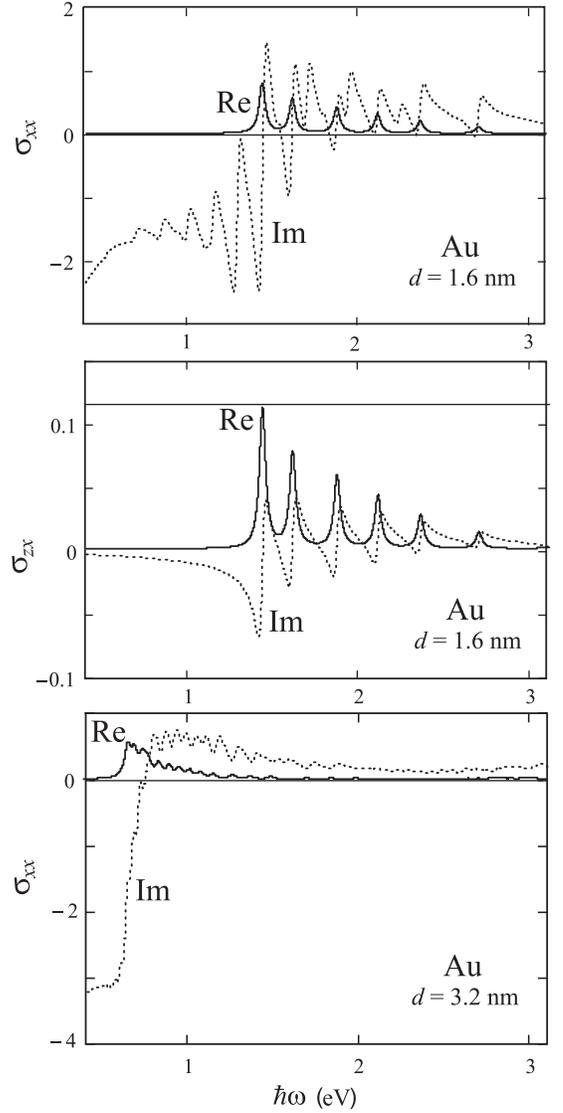}
\caption{Calculated frequency dependences of the real and imaginary
parts of  $\sigma_{\alpha\beta}$ (in $e^{2}/2a_{0}\hbar$ units) for
Au wires of various diameter $d$.} \label{KP-fig7}
\end{figure}

Let us find, for example, the position of the peak in ${\rm
Re}\,\sigma_{xx}$ which has the maximum height. The height of the
peaks is proportional to $\left|\langle
m+1,n'|\hat{p}_{x}|mn\rangle\right|^{2}\sum_{p}1$. The matrix
element has the maximum magnitude at $n'=n$ because under this
condition the integral in (\ref{KKBP-13}) takes on the maximum
value. Further, by using (\ref{KKBP-2009}) and (\ref{KKBP-2010}), it
is easy to determine that the maximum height is realized at $m=0$,
$n'=n=n_{\rm F}$. For the diameter $d=1.6$ nm, the number $ n_{\rm
F}\approx k_{\rm F}^{0}\rho_{0}/\pi$ is equal to 3. As a result we
have
$$
\hbar\omega_{\rm
max}=\frac{\hbar^{2}\left(k_{1,3}^{2}-k_{0,3}^{2}\right)}{2m_{e}}=
\frac{\hbar^{2}\left(a_{1,3}^{2}-a_{0,3}^{2}\right)}{2m_{e}\rho_{0}^{2}}=1.40\,\
\rm eV.
$$
This is in good agreement with numerical calculations presented in
Fig. 7. The height of the peaks in Fig. 7 also confirms our
estimation.

Fig. 7 demonstrates the important fact  ${\rm Re}\,\sigma_{xx}>0$
over all frequency range. In contrast to this, ${\rm
Im}\,\sigma_{xx}$ is a variable in sign function of the frequency.

The frequency dependences of ${\rm Re}\,\sigma_{zx}$ and ${\rm
Im}\,\sigma_{zx}$ are also presented in Fig. 7. As it was expected,
the position of the peaks is identical both for ${\rm
Re}\,\sigma_{zx}$ and ${\rm Im}\,\sigma_{zx}$ but the height is one
order smaller in the first case.

Comparing the upper and lower parts of Fig. 7, we can trace the size
dependence of the conductivity for ultrathin metal wires. When $d$
increases, the peaks shift to the left with displacement equal to
$\Delta\omega=\omega'-\omega=\omega(\rho_{0}^{2}/{\rho'_{0}}^{2}-1)$.
More distant peaks (with lager value $\omega$) have lager
displacement, so that the interval occupied by the peaks contracts.
At the same time, new peaks appear within this interval, because
with enlarging $\rho_{0}$ the number of levels and the number of the
possible transitions between them increase. Distance between peaks
decreases, and when it approaches $\hbar/\tau$, the peaks begin to
merge together.

It is interesting to compare results of the study for the optical
conductivity of ultrathin metal wires with analogous results for
ultrathin films. Divergences are associated with the different
dimensionality of the systems. This is reflected in an essential
difference in the energetic spectra and also in the fact that after
calculation of quasicontinuous states, in the case of a wire, the
summation over two numbers $m, n$ remains, while in the case of a
film, it remains over one number only (which numerates values of the
$x-$component of the electron momentum). It is this fact that
explains approximately one order lower height of the maximum in the
frequency dependence of the conductivity of a wire compared to the
case of a film. Indeed,
$$
\frac{{\rm Re}\,\sigma_{xx}^{\rm wire}}{{\rm Re}\,\sigma_{xx}^{\rm
film}}\cong\frac{\Omega_{\rm wire}^{-1}\sum_{p}1}{\Omega_{\rm
film}^{-1}\sum_{p,n}1}=\frac{2{\mathcal L}}{\pi}\frac{\sqrt{(k_{\rm
F}^{\rm wire})^{2}-k_{0n_{\rm F}}^{2}}}{\pi\rho_{0}^{2}{\mathcal L}}
$$
$$
\times\left\{\frac{ab}{\pi^{2}}\frac{\pi\{(k_{\rm F}^{\rm
film})^{2}-k_{m{\rm F}}^{2}\}}{abL}\right\}^{-1}\cong
\frac{10^{-1}}{\sqrt{\rho_{0}[{\rm nm}]}}\frac{L^{2}}{\rho_{0}^{2}}
$$
because
$$
(k_{\rm F}^{\rm wire})^{2}-k_{0n_{\rm F}}^{2}\cong 2\pi k_{\rm
F}^{0}/\rho_{0},\,\,\,(k_{\rm F}^{\rm film})^{2}-k_{m_{\rm
F}}^{2}\cong 2\pi k_{\rm F}^{0}/L.
$$
For   $L,\rho_{0}\cong 1$ nm we obtain ${\rm Re}\,\sigma_{xx}^{\rm
wire}/{\rm Re}\,\sigma_{xx}^{\rm film}\cong10^{-1}$.

As to the different position of the peaks, this may be completely
explained by characteristic properties of spectra of the 1D and 2D
systems.

The frequency dependences of ${\rm Re}\,\sigma_{xx}$ for the Al and
Pb wires of diameter 1.6 nm are presented in Fig. 8. It is
surprising that peaks in the conductivity of the Pb wire are
absolutely absent. The reason of this is a small value of the
relaxation time for Pb equal to $\tau=1.4\times 10^{-15}$ s, so that
width of the peaks $\hbar/\tau=0.44$ eV. In this respect, Al, with
$\hbar/\tau=0.08$ eV, holds an intermediate position between Au and
Pb. For calculations we use values of the relaxation time for bulk
metals taken from \cite{27}.
\begin{figure}[!t!b!p]
\centering
\includegraphics [width=.4\textwidth] {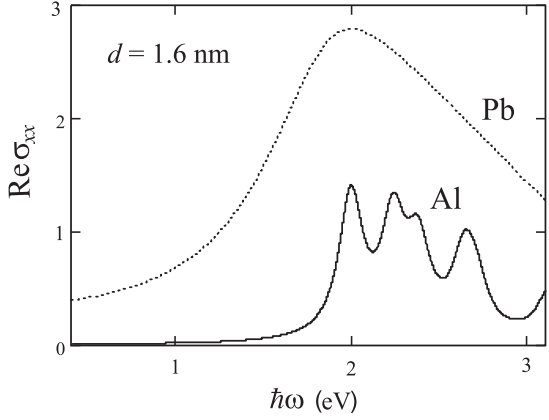}
\caption{Calculated frequency dependences of the real part of
$\sigma _{xx}$ (in $e^{2}/2a_{0}\hbar$ units) for Al and Pb wires.}
\label{KP-fig8}
\end{figure}

In spite of an absence of peaks in the frequency dependence of the
conductivity for the Pb wire, its maximum may be found in such a way
as the position of the maximum height peak in the conductivity of
the Au wire was determined above, with the difference that this time
$n_{\rm F}=4$:
$$
\hbar\omega_{\rm
max}=\frac{\hbar^{2}\left(a_{1,4}^{2}-a_{0,4}^{2}\right)}{2m_{e}\rho_{0}^{2}}=2.1\,\
\rm eV.
$$
This value agrees well with Fig. 8 taking into account the large
width of the peaks.

Surprisingly, the difference in the obtained results for Al and Pb
(due to the different values of  $\tau$) indicate size-frequency
dependences, which can be expected for films and wires,
inhomogeneous in thickness considered, for example, in Refs.
\cite{Tesanovic,d19,d20,d21,d22}.  If fluctuations of sizes in 1D-
and 2D-systems lead to a strong effective reduction of $\tau$,
experimental size dependences of conductivity are noticeably
smoothed irrespective of a metal kind.

We will devote select publication to the theory of transport in
films and wires with rough surface.

\section{Conclusions}

The conductivity tensor is introduced for the low-dimensional
electron systems. Components of the conductivity tensor for a
quasi-homogeneous ultrathin metal film and wire are calculated
within the particle-in-a-box model on the assumption that the
component of the induced current with the wave vector equal to the
wave vector of the electromagnetic field is dominating.

Over infrared region the condition  $qL,\,q\rho_{0}\ll 1$  is
satisfied allowing us to express components of the conductivity
tensor in terms of the according small value. All non-diagonal
components of the conductivity tensor are equal to zero in zero
order of the expansion. They appear in linear approximation. The
important fact that the real part of the diagonal components is
non-negative over all frequency range, with the guarantee $Q > 0$
for the dissipation of energy, is proved.

As a result of comparing the according components of the
conductivity tensor for a film and a wire of the same thickness of
order 1 nm, one order smaller value for a wire is obtained. In such
a manner different density of states near the Fermi level manifests
itself (it is greater for a film). It is found that the discrepancy
between our results and the theory \cite{18} increases and becomes
substantial, as the characteristic small dimension of the system
decreases. This discrepancy is associated with the strong dependence
of the Fermi level on this dimension for small values of order of
the Fermi wavelength. This size dependence of the Fermi level has an
``oscillatory'' form. Transmittance is calculated for a simple,
 well-defined model without fitting parameters.

\acknowledgments {We are grateful to A. V. Babich and  A. V. Korotun
for help in the calculations and to W. V. Pogosov for reading the
manuscript. This work was supported by the Ministry of Education and
Science of Ukraine.}

\begin{appendix}

\section{The matrix elements}

The expressions for momentum projections in cylindrical coordinates
have a form
$$
\hat p_z  =  - i_{0}\hbar \frac{\partial }{{\partial z}},
$$
\begin{equation}
\hat p_x = - i_{0}\hbar \left\{ {\cos \varphi \frac{\partial
}{{\partial \rho }} - \frac{{\sin \varphi }}{\rho }\frac{\partial
}{{\partial \varphi }}} \right\},\label{KKBP-A1}
\end{equation}
$$
\hat p_y = - i_{0}\hbar \left\{ {\sin \varphi \frac{\partial
}{{\partial \rho }} + \frac{{\cos \varphi }}{\rho }\frac{\partial
}{{\partial \varphi }}} \right\}.
$$

Using (\ref{PPP}) -- (\ref{KKBP-5a})  and (\ref{KKBP-A1}), we have

\begin{widetext}
\begin{multline}
\left\langle j\right|\hat p_{x} \left|i\right\rangle
=-i_{0}\frac{\hbar}{2}
\delta_{pp'}\left\{\frac{1}{2\pi}\int\limits_0^{2\pi}\left(e^{-im'\varphi}e^{i(m-1)\varphi}+e^{-im'\varphi}e^{i(m+1)\varphi}\right)d\varphi
\int\limits_0^{\rho_{0}}R_{m'n'}\frac{dR_{mn}}{d\rho}\rho
d\rho\right.
\\
+\left.\frac{m}{2\pi}
\int\limits_0^{2\pi}\left(e^{-im'\varphi}e^{i(m-1)\varphi}-e^{-im'\varphi}e^{i(m+1)\varphi}\right)d\varphi
\int\limits_0^{\rho_{0}}R_{m'n'}R_{mn}d\rho\right\}
\\
=-i_{0}\frac{\hbar}{2}\delta_{pp'}\left\{\delta_{m-1,m'}\left(
\int\limits_0^{\rho_{0}}R_{m'n'}\frac{dR_{mn}}{d\rho}\rho d\rho+
m\int\limits_0^{\rho_{0}}R_{m'n'}R_{mn}d\rho\right)\right.
\\
+ \left.\delta_{m+1,m'}\left(
\int\limits_0^{\rho_{0}}R_{m'n'}\frac{dR_{mn}}{d\rho}\rho d\rho-
m\int\limits_0^{\rho_{0}}R_{m'n'}R_{mn}d\rho\right)\right\};
\label{KKBP-A2}
\end{multline}
\end{widetext}
Then using relation  $ I'_{m}(x)=\pm mI_{m}(x)/x \mp I_{m\pm 1}(x)$
from \cite{Yanke}, we obtain Eqs. (\ref{KKBP-12}) --
(\ref{KKBP-13}).

In a similar way we find
\begin{widetext}
\begin{multline}
\left\langle j\right|x\hat p_{z}\left|i\right\rangle = -i_{0}\hbar
\int\int\int\left\{R_{m'n'}\Phi_{m'}^{*}Z_{p'}^{*}(\rho\cos\varphi)
R_{mn}\Phi_{m}\frac{dZ_{p}}{dz}\right\}\rho d\rho d\varphi dz
\\
=\hbar
k_{zp}\delta_{pp'}\int\limits_0^{2\pi}\Phi_{m'}^{*}\Phi_{m}\cos\varphi
d\varphi \int\limits_0^{\rho_{0}} R_{m'n'}R_{mn}\rho^{2}d\rho
=\frac{1}{2}\hbar
k_{zp}\delta_{pp'}(\delta_{m-1,m'}+\delta_{m+1,m'})
\int\limits_0^{\rho_{0}} R_{m'n'}R_{mn}\rho^{2}d\rho
\\
=\frac{1}{2}\hbar k_{zp}\delta_{pp'}
\left(\delta_{m-1,m'}\int\limits_0^{\rho_{0}}R_{m-1,n'}R_{mn}\rho^{2}d\rho
+\delta_{m+1,m'}\int\limits_0^{\rho_{0}}R_{m+1,n'}R_{mn}\rho^{2}d\rho\right).
\label{KKBP-B3}
\end{multline}
\end{widetext}

\end{appendix}

\end {document}